\documentclass[10pt,twocolumn,twoside,letterpaper]{IEEEtran}

\usepackage{geometry}
\makeatletter
\def\ps@IEEEtitlepagestyle{
  \def\@oddfoot{\mycopyrightnotice}
  \def\@evenfoot{}
}
\def\mycopyrightnotice{
  {\footnotesize
  \begin{minipage}{\textwidth}
  \centering
  978-1-7281-4164-0/19/\$31.00 \copyright2019 IEEE
  \end{minipage}
  }
}

\usepackage{url}
\usepackage{graphicx}

\begin{document}
%
\title{A 1 m$^3$ Gas Time Projection Chamber with Optical Readout for Directional Dark Matter Searches: the CYGNO Experiment}
%
%
%

\author{E. Baracchini, R. Bedogni, F. Bellini, L. Benussi, S. Bianco, C. Capoccia, M. Caponero, G. Cavoto, I. A. Costa, E. Di Marco, G. D'Imperio, F. Iacoangeli, G. Maccarone, M. Marafini, G. Mazzitelli, A. Messina, A. Orlandi, E. Paoletti, L. Passamonti, A. Pelosi, F. Petrucci, D. Piccolo, D. Pierluigi, D. Pinci, F. Renga, A. Russo, G. Saviano and S. Tomassini
\thanks{Manuscript received December 13, 2019. This work was supported in part by the European Research Council (ERC) under the European Union's Horizon 2020 research and innovation programme (grant agreement No 818744).}
\thanks{I. A. Costa, E. Di Marco, G. D'Imperio, F. Iacoangeli, A. Pelosi, D. Pinci and F. Renga (e-mail: francesco.renga@roma1.infn.it) are with 
Istituto Nazionale di Fisica Nucleare, Sezione di Roma, Roma, Italy.}
\thanks{E. Baracchini is with Gran Sasso Science Institute, L'Aquila, Italy and Istituto Nazionale di Fisica Nucleare, Laboratori Nazionali del Gran Sasso, Assergi, Italy.}
\thanks{R. Bedogni, L. Benussi, S. Bianco, C. Capoccia, G. Maccarone, G. Mazzitelli, A. Orlandi, E. Paoletti, L. Passamonti, D. Piccolo, D. Pierluigi A. Russo and S. Tommasini are with Istituto Nazionale di Fisica Nucleare,  Laboratori Nazionali di Frascati, Frascati,  Italy.}
\thanks{F. Bellini, G. Cavoto and A. Messina are with Dipartimento di Fisica,  Sapienza Universit\`a di Roma and Istituto Nazionale di Fisica Nucleare, Sezione di Roma, Roma, Italy.}
\thanks{M. Caponero is with ENEA Centro Ricerche Frascati and Istituto Nazionale di Fisica Nucleare, Laboratori Nazionali di Frascati, Frascati,  Italy.}
\thanks{M. Marafini is with Museo Storico della Fisica e Centro Studi e Ricerche "Enrico Fermi" and Istituto Nazionale di Fisica Nucleare, Sezione di Roma, Roma, Italy.}
\thanks{F. Petrucci is with Dipartimento di Matematica e Fisica, Universit\`a Roma Tre and Istituto Nazionale di Fisica Nucleare, Sezione di Roma Tre, Roma, Italy}
\thanks{G. Saviano is with Dipartimento di Ingegneria Chimica, Materiali e Ambiente, Sapienza Universit\`a di Roma, Roma, Italy and Istituto Nazionale di Fisica Nucleare,  Laboratori Nazionali di Frascati, Frascati,  Italy}}

\maketitle

\pagenumbering{gobble}

\begin{abstract}
The aim of the CYGNO project is the construction and operation of a 1~m$^3$ gas TPC for directional dark matter searches
and coherent neutrino scattering measurements, as a prototype toward the 100-1000~m$^3$ (0.15-1.5 tons) CYGNUS network
of underground experiments. In such a TPC, electrons produced by dark-matter- or neutrino-induced nuclear recoils will drift 
toward and will be multiplied by a three-layer GEM structure, and the light produced in the avalanche processes will be readout by a 
sCMOS camera, providing a 2D image of the event with a resolution of a few hundred micrometers. Photomultipliers will also provide 
a simultaneous fast readout of the time profile of the light production, giving information about the third coordinate and hence allowing 
a 3D reconstruction of the event, from which the direction of the nuclear recoil and consequently	 the direction of the incoming particle can be 
inferred. Such a detailed reconstruction of the event topology will also allow a pure and efficient signal to background discrimination. 
These two features are the key to reach and overcome the solar neutrino background that will ultimately limit non-directional dark matter searches.
\end{abstract}


\vspace{2 cm}

\section{Introduction}
%
%
%
%

\IEEEPARstart{T}{he} search for dark matter is one of the main topics of experimental particle physics in these years. Most of the experiments
are performed following the well known WIMP paradigm: dark matter could be made of weakly-interacting particles (with a cross section for interaction
with standard matter around or below $10^{-40}~\mathrm{cm}^2$) with a mass in the GeV to TeV range. These assumptions match indeed the astrophysical and 
cosmological evidences for dark matter and the predictions for the lightest stable new particle in many supersymmetric models, making this scenario
very attractive.

\begin{figure}[!b]
\centering
\includegraphics[width=\linewidth, angle=0]{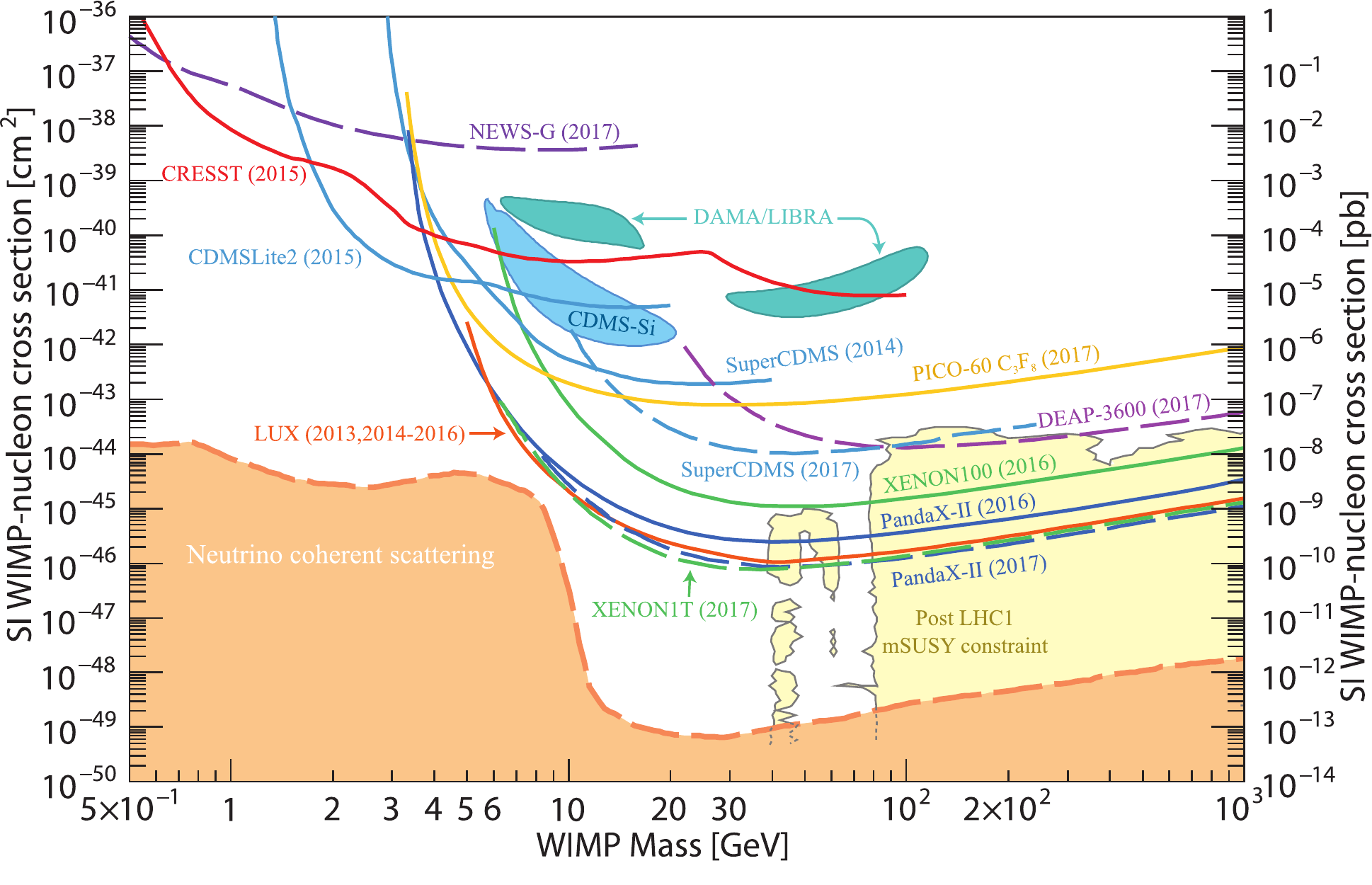}
\caption{Current limits on WIMP-like dark-matter spin-independent inrteractions, from~\cite{pdg}.}
\label{fig:DM-limits}
\end{figure}

The very small cross section for interaction with standard matter pushed the experimental efforts toward the development of large and high-density detectors,
with LXe two-phase time projection chambers (TPCs) leading the field, see Fig.~\ref{fig:DM-limits}. On the other hand, some limitations of such 
an approach are starting to emerge:
\begin{itemize}
\item Large-mass target nuclei imply a very low nuclear recoil energy, below the detecting threshold, if the mass of the WIMP is small. It makes the
region around and below 1~GeV almost unexplored so far.
\item The sensitivity of the next-generation experiments will approach the so-called neutrino floor, i.e. the level at which the detectors will start
to collect events of recoils from coherent scattering of solar, atmospheric and supernova neutrinos on nuclei. These events will be undistinguishable 
from dark-matter-induced recoils and hence would constitute an irreducible background for dark matter searches.
\end{itemize}
Moreover, the striking evidence for an annual modulation in the rate of nuclear recoils in the DAMA experiment~\cite{dama} and all controversies around it
demonstrate that some peculiar feature on data is needed to claim unambiguously the observation of a dark matter signal.

A worldwide effort is ongoing to overcome these difficulties by developing gas TPCs with Helium-based mixtures for dark matter searches. Helium
would provide a light target, experiencing recoils with relatively high energy (a few keV or more) even from scattering of WIMPs of GeV or sub-GeV mass. 
The low density of the gas would also make such recoiling nuclei to travel millimetric distances, making it possible to reconstruct the direction of
the recoil. It would give in turn an information about the direction of the incoming particle. Since the earth moves in the dark matter halo of our galaxy, 
and the event rate is expected to be larger for particles seen in the direction of the earth motion (corresponding to the direction of the Cygnus constellation),
the directional capabilities of such detectors would allow to distinguish a dark matter signal from backgrounds generally coming from other directions.
The drawback of such a low-mass target is of course the large volume that is needed to reach a mass large enough to be competitive in terms of exposure
with experiments made of solid or liquid targets. Anyway, services needed by a gaseous detector at atmospheric pressure are typically compact and the cost 
per unit volume relatively small, so that active volumes of 100~m$^3$ (corresponding to 100~kg with the gas mixtures under study) are not impossible to 
assemble is a single site, while the ton scale could be reached by a network of similar detectors in multiple sites, as envisaged by the worldwide collaboration
named CYGNUS-TPC.

We discuss here the CYGNO experiment, a project for the development of a 1~m$^3$ TPC instrumented with optically readout gas electron multipliers (GEMs),
as a part of the CYGNUS-TPC effort. Such a detector would work as a demonstrator for larger scale experiments based on the same technology, providing 
at the same time interesting limits in the GeV WIMP mass region and opening other interesting opportunities in neutrino physics.

We also introduce the INITIUM project, aiming at the development of gas TPCs with negative ion drift, which would allow to overcome some of the limitations
of the conventional gas TPC approach.

\section{The GEM optical readout}

Gas electron multipliers are widely used since several years as the multiplication stage in gaseous TPCs, mainly thanks to their stability at high rates
and their capability of suppressing the back-flow of positive ions produced in the avalanche toward the drift volume. Although these requirements are not
critical in a dark matter detector, GEMs turned out to be a good choice in this field for other reasons we will illustrate here, if the optical readout 
technique is adopted.

In electron avalanches many photons are usually produced, mostly in the UV range but also in the visible region. The yield of visible light is particularly enhanced in
gas mixtures containing scintillating components like CF$_4$. If pictures of a GEM in the dark are taken, avalanches can be seen as light spots. It allows a 2-dimensional 
reconstruction of events occurring in the detector, with a very high granularity at a reasonable cost, considering that a single photo camera provides millions of
pixels and can be used to image an arbitrarily large surface. The limits are imposed only by the amount of light to be collected to have a sufficient signal over noise ratio.
It makes this technique very attractive for directional dark matter searches, that require very large detectors with a granularity high enough to reconstruct 
submillimetric tracks.

\begin{figure}[!t]
\centering
\includegraphics[width=\linewidth, angle=0]{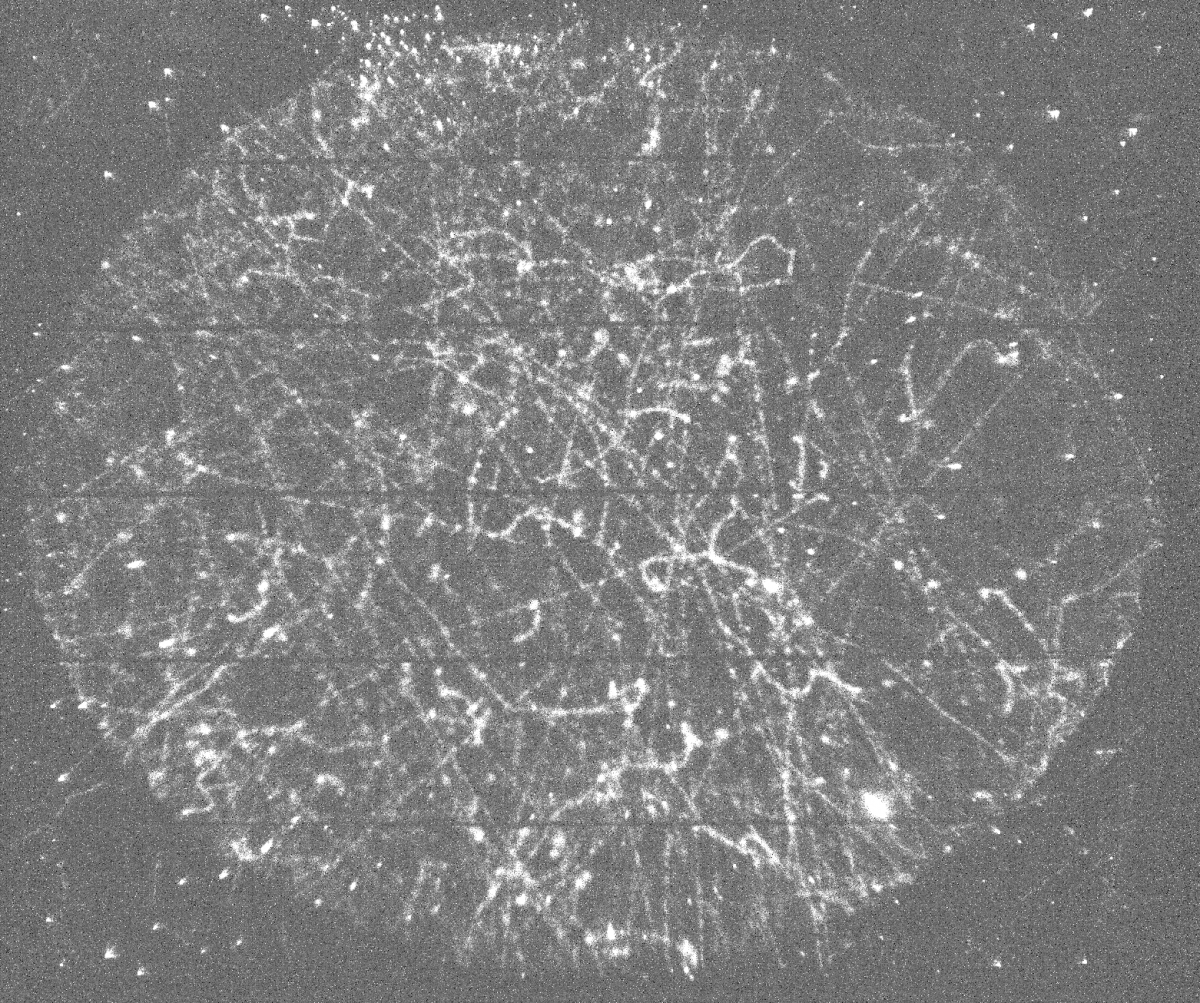}
\caption{Picture of 5 second exposure of sCMOS camera to a 20~cm $\times$ 24~cm triple-GEM detector with 30~cm drift field.}
\label{fig:lemon-pic}
\end{figure}

The optical readout of GEMs with sCMOS cameras was successfully investigated in the last decade~\cite{optical}. He:CF$_4$ mixtures were studied, and
volume compositions from 60:40 to 70:30 where found to be the most luminous and stable. A picture of a 20~cm $\times$ 24~cm triple-GEM detector with 20~cm
drift volume, imaged by an Hamamatsu Orca Flash 4.0 camera~\cite{orca}, with a 25~mm focal length, f/0.95 aperture optics and an exposure of 5 seconds is shown in 
Fig.~\ref{fig:lemon-pic}. The camera was placed at a distance of about 60~cm from the GEM, so that the full surface of the GEM could be framed. Different species of 
events can be recognized, including cosmic rays and different kinds of natural radioactivity.

The light produced by avalanches in GEMs can be also captured by photon detectors like PMTs or SiPMs. In this case, the position information is lost but the very fast 
response allows to study the time development of the track. In a TPC, where the track can travel a significative distance along the drift field, electrons produced along
the track can arrive at different times that can be resolved by fast photon detectors (typical drift velocities being in the order of 30~$\mu$m/ns). This information
can be exploited to study the development of the track along the drift direction. In combination with the sCMOS picture (\emph{double optical readout}), 
it provides a 3-dimensional reconstruction of the track~\cite{combined}.

\section{The CYGNO project}

The aim of the CYGNO project is the construction of a 1~m$^3$ gas TPC with a He:CF$_4$ mixture, that is intended as a demonstrator for a 100~kg-scale detector
for the directional search of low-mass WIMPs, to be installed at Laboratori Nazionali del Gran Sasso (LNGS), Italy. Considering that the density of the gas mixtures under 
study is typically $O(1~\mathrm{kg}/\mathrm{m}^3)$, the CYGNO detector will have an active mass of about 1~kg.

\begin{figure}[!t]
\centering
\includegraphics[width=\linewidth, angle=0]{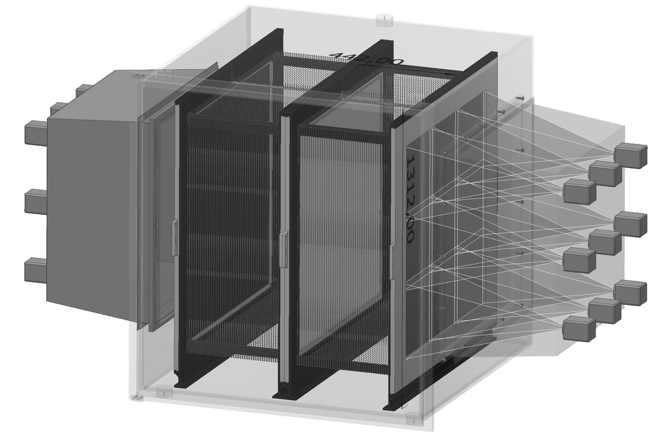}
\caption{Conceptual design of the CYGNO experiment.}
\label{fig:cygno}
\end{figure}

The detector will be made of 2 face-to-face drift volumes, 50~cm long, separated by a cathode and readout at the two sides by triple-GEM stacks, 1~m$^2$ total surface each. A set of 18 cameras (9 per side) will be used, framing a 33~cm $\times$ 33~cm surface each. A set of PMTs or SiPMs will be also installed in between the 
cameras, also looking at the GEMs, to allow the reconstruction of the third coordinate. A conceptual design of the detector is shown in Fig.~\ref{fig:cygno}.

The design of the detector will be completed in 2020 and followed by the construction, with the goal of running the experiment at LNGS starting around 2021.

\section{The CYGNO prototypes}

A few prototypes are used to assess the performances of the technologies to be used in CYGNO. The first tests of the double optical readout have been performed
with a small prototype (\emph{ORANGE}) equipped with a 10~cm $\times$ 10~cm triple-GEM and a drift volume of 1~cm depth.

\begin{figure}[!b]
\centering
\includegraphics[width=\linewidth, angle=0]{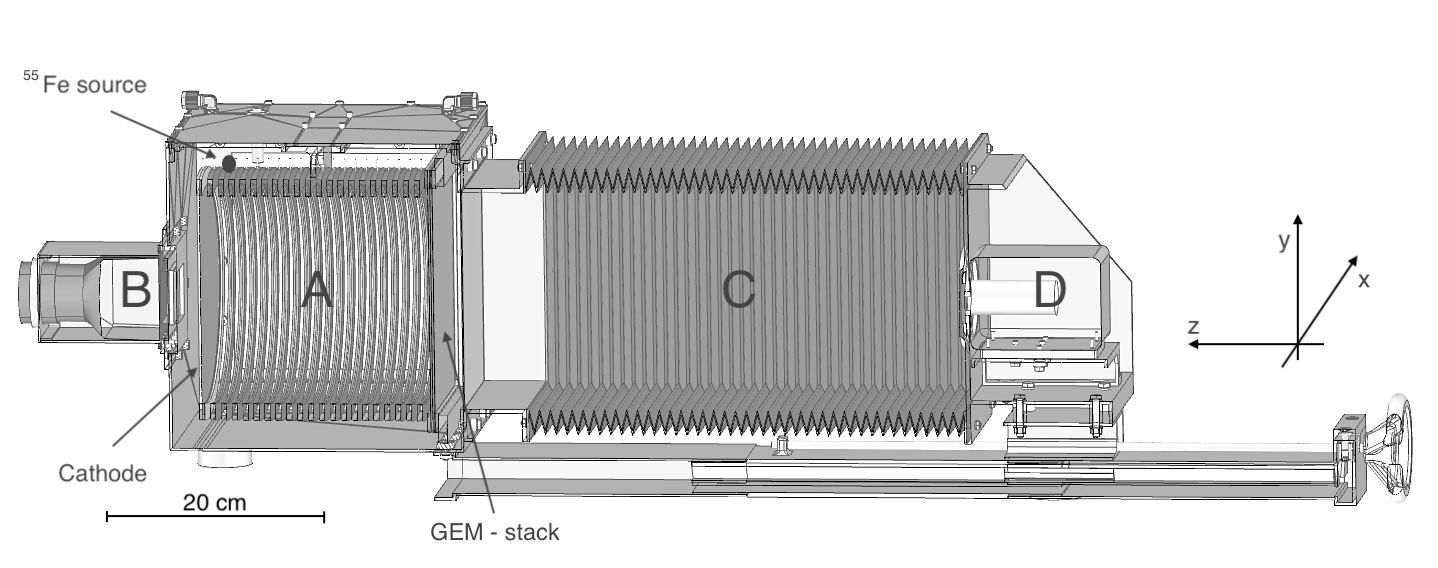}
\caption{The LEMON prototype, with the field cage (A), the PMT (B), the bellow to change the position of the photo camera (C), and the camera stage (D).}
\label{fig:lemon}
\end{figure}

A second, larger prototype was built (\emph{LEMON}) with 20~cm $\times$ 24~cm triple-GEM and 20~cm drift volume (see Fig.~\ref{fig:lemon}). 
The uniformity of the field in the drift volume is guaranteed by a field cage made of a set of metal rings embedded in 3D-printed holders and kept at a voltage 
linearly increasing with the distance from the GEM stack. The field cage is completed by a cathode made of a fine wire mesh. The transparency of the cathode
allowed to place the PMT behind it. Drift fields up to several hundred V/cm were tested.

Currently, a third prototype (\emph{LIME}) is under construction, with the aim of testing the main construction parameters of CYGNO. The sCMOS camera will frame
a surface of 33~cm $\times$ 33~cm from a distance of about 60~cm. The drift volume will be 50~cm deep. Different field cages and cathodes will be tested, 
including a conventional set of metal electrodes or a continuous resistive sheet, providing the correct voltage gradient. The gas-tight container will be built in PMMA,
as we foresee to do in CYGNO.

\section{Detector performances and physics reach}

\begin{figure}[!b]
\centering
\includegraphics[width=\linewidth, angle=0]{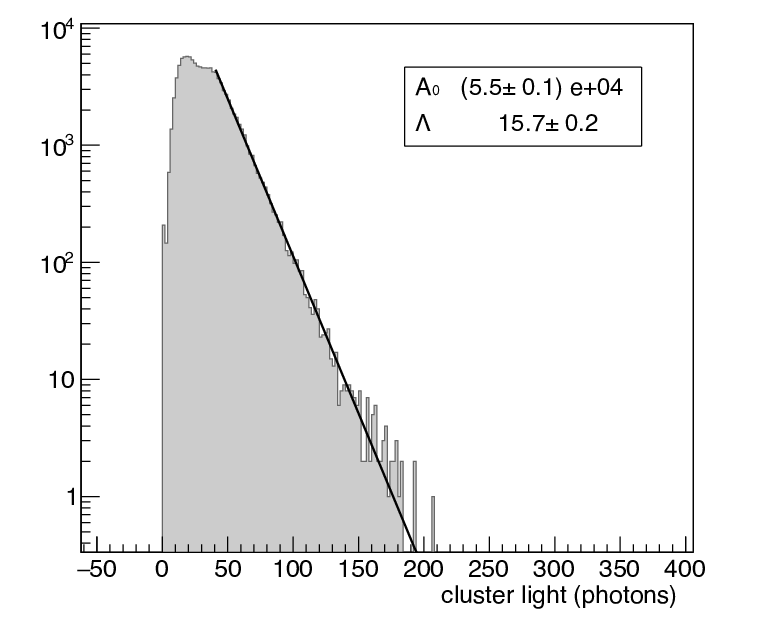}
\caption{Distribution of the light in clusters reconstructed in a run with blind sensor.}
\label{fig:noise}
\end{figure}

The main specifications of the sCMOS camera are quoted in Table~\ref{tab:camera}. We performed a measurement of the sCMOS sensor noise taking data with the 
camera completely blinded~\cite{lemon}. Contiguous illuminated pixels are grouped in clusters, and the distribution of the total light in the clusters is shown in 
Fig.~\ref{fig:noise}. From this plot we expect $< 10$ events per year if a threshold of 400 photons is set, without any other kind of selection on the cluster shape.
In this setup, 400 photons correspond to about 2~keV of energy released in the detector. It indicates that, from the instrumental point of view, 
a threshold of a few keV can be easily reached.

\begin{table}[h!]
\caption{\label{tab:camera} Specification of the Hamamatsu Orca Flash 4.0 camera}
\begin{center}
\begin{tabular}{cc}
\hline
\hline
Pixels & 2048 x 2048 \\
Pixel size & 6.5 $\mu$m \\
Effective area & 13.312 $\times$ 13.312~mm$^{2}$ \\
Readout noise & 1.4e RMS\\
Resolution & 16 bit\\
\hline
\hline
\end{tabular}
\end{center}
\label{default}
\end{table}%

\begin{figure}[!t]
\centering
\includegraphics[width=\linewidth, angle=0]{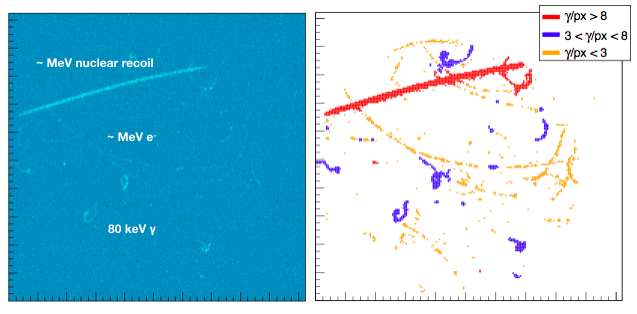}
\caption{An example of cluster reconstruction and particle identification with the ORANGE prototype. The raw picture (left) is processed to reconstruct clusters 
and assign them to different species of events (right).}
\label{fig:clusters}
\end{figure}

On the other hand, the experimental threshold of the detector, when installed underground at LNGS, will be eventually determined by the capability 
of rejecting background from natural radioactivity. We started a detailed simulation of the background expected in CYGNO, taking into account the 
radioactivity spectra measured at LNGS and including different options
for the detector shielding. Preliminary results indicate that a $\gamma$ rejection capability of $O(10^5 - 10^6)$ is required. It can be reached by exploiting the
very detailed characterization of the track shape provided by the high granularity of the optical readout. Fig.~\ref{fig:clusters}, which was taken with the ORANGE
prototype, shows how it is possible to separate and identify clusters produced by different kinds of natural radioactivity. Activities to develop pattern recognition 
algorithms to automatically identify them are ongoing.

Another critical aspect of the experiment is the capability of inferring the direction of the nuclear recoil. Recent tests show that the
shape of recoil tracks down to 15~keV is still asymmetric enough to extract a directional information. The determination of the resolution
that can be reached for such a low energy will be the objective of studies in the next future.

The main limiting factor for both background rejection and directionality is the diffusion of the electrons drifting in the gas. Diffusion in the He:CF$_4$ (60:40) gas
mixture was studied in LEMON by measuring the reconstructed size of clusters produced by a collimated $^{55}$Fe X-ray source as a function of its position along 
the drift volume~\cite{diffusion}. Thanks to the very low diffusion in CF$_4$, a diffusion coefficient of only $\sim 130~\mu\mathrm{m}/\sqrt{\mathrm{cm}}$ was measured.

Diffusion provides on the other hand the possibility of inferring the position of the track along the drift direction. Indeed, the absence of a prompt signal 
associated to the track prevents to reconstruct this position from a drift time measurement. Instead, the smearing of the reconstructed track due to the diffusion 
and its dependence on the drift distance allow to distinguish track produced at different distances from the GEMs. This possibility was investigated using 
$\sim 450$~MeV electron tracks at the Beam Test Facility (BTF) of INFN Laboratori di Frascati. The size of the observed clusters, normalized by their amplitude, is
influenced by the diffusion, that smears more clusters produced farther from the GEMs. It allows to reconstruct the coordinate along the drift direction with a resolution 
of about 7\%, corresponding to 3.5~mm at the maximum drift distance in CYGNO (50~cm). This feature is crucial to reject events happening near the GEMs or the 
cathode, which could be due to the radioactivity of the materials, and hence define a fiducial volume (\emph{fiducialization}) with a minimum loss of active mass.

\begin{figure}[!h]
\centering
\includegraphics[width=0.8\linewidth, angle=0]{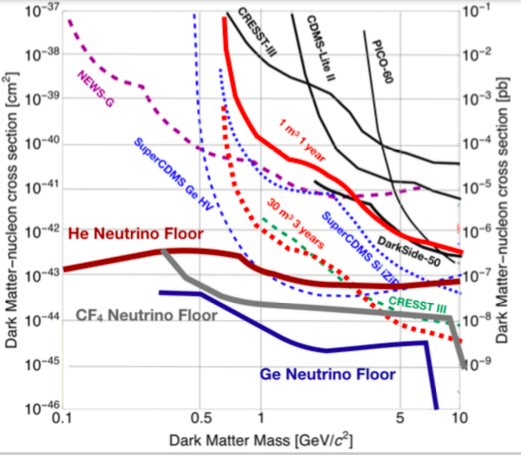}
\caption{Expected limits from the CYGNO experiment (1-year exposure of 1~m$^3$ detector) and from a 3-year exposure of a 30~m$^3$ detector.}
\label{fig:limits-cygno}
\end{figure}

Putting together our current estimates of the detector performances, and considering either a 1-year exposure of a 1~m$^3$ detector or a 3-year exposure of a 30~m$^3$
detector, we expect to set the limits shown in Fig.~\ref{fig:limits-cygno}.

\section{The negative-ion option: the INITIUM project}

Two of the main limitations of the technique proposed for CYGNO are the impact of diffusion on the directional and background rejection capabilities, and the 
difficulties connected to the fiducialization based on diffusion. An idea to overcome these issues consists in adding to the gas mixture a small amount of an
extremely electronegative gas like SF$_6$. When free electrons are produced by ionization, they can be captured by the electronegative gas to form different species
of negative ions like SF$_6^-$ or SF$_5^-$. Negative ions would drift toward the GEM stack, where they can release the electron due to
the strong electric field and produce a normal electron avalanche. The main advantages of this technique rely on the very low diffusion experienced by negative
ions and on the presence of different ion species. The low diffusion coefficient would produce very good images of the event. The different species would drift at
different velocities, and arrive at the GEMs with time delays proportional to the drift distance. It would allow a precise measurement of the coordinate along the 
drift direction.

While previous attempts of using negative-ion drift have been made with pure electronegative gases at low pressure~\cite{drift}, in 2016 it has been demonstrated 
for the first time that negative-ion operation is possible in a more conventional gas mixture at nearly atmospheric pressure with the addition of a small 
SF$_6$ component~\cite{nitec}. It opened the way to the combination of negative ion drift and optical readout.

The goal of the INITIUM project is to demonstrate that negative-ion operation can be advantageous in a setup similar to the one of the CYGNO experiment. The
construction of a small prototype (\emph{MANGO}), to be operated in these conditions, with 10~cm $\times$ 10~cm GEMs and 5~cm drift region, is currently 
ongoing and will be followed by tests of negative-ion operation in CYGNO.





\begin{thebibliography}{1}

\bibitem{pdg}
  M.~Tanabashi {\it et al.} [Particle Data Group],
  \emph{Review of Particle Physics},
  Phys.\ Rev.\ D {\bf 98}, no. 3, 030001 (2018).

\bibitem{dama}
  R.~Bernabei {\it et al.} [DAMA and LIBRA Collaborations],
  \emph{New results from DAMA/LIBRA},
  Eur.\ Phys.\ J.\ C {\bf 67}, 39 (2010)

\bibitem{optical} 
  M.~Marafini, V.~Patera, D.~Pinci, A.~Sarti, A.~Sciubba and E.~Spiriti,
  \emph{Optical readout of a triple-GEM detector by means of a CMOS sensor},
  Nucl.\ Instrum.\ Meth.\ A {\bf 824}, 562 (2016).

\bibitem{orca}
\url{https://www.hamamatsu.com/eu/en/product/type/C13440-20CU/index.html}

\bibitem{combined} 
  V.~C.~Antochi {\it et al.},
  \emph{Combined readout of a triple-GEM detector},
  JINST {\bf 13}, no. 05, P05001 (2018).

\bibitem{lemon}
 I.~Abritta~Costa {\it et al.}, 
 \emph{Performance of optically readout GEM-based TPC with a $^{55}$Fe source}
 JINST {\bf 14}, no. 7, P07011 (2019).

\bibitem{diffusion} 
  G.~Mazzitelli {\it et al.},
  \emph{MPGD Optical Read Out for Directional Dark Matter Search},
  2018 IEEE Nuclear Science Symposium and Medical Imaging Conference Proceedings (NSS/MIC), Sydney, Australia, 2018, pp. 1-4.

\bibitem{drift} 
  G.~J.~Alner {\it et al.},
  \emph{The DRIFT-I dark matter detector at Boulby: Design, installation and operation},
  Nucl.\ Instrum.\ Meth.\ A {\bf 535}, 644 (2004).

\bibitem{nitec}
  E.~Baracchini, G.~Cavoto, G.~Mazzitelli, F.~Murtas, F.~Renga and S.~Tomassini,
  \emph{Negative Ion Time Projection Chamber operation with SF$_6$ at nearly atmospheric pressure},
  JINST {\bf 13}, no. 04, P04022 (2018).


\end{thebibliography}
%




\end{document}